\documentclass[12pt]{article}
\usepackage{amsfonts}
\usepackage{graphicx}
\usepackage{latexsym}
\usepackage{epsfig}
\usepackage{amsmath}
\usepackage{amssymb}
\usepackage{color}
\usepackage{subfigure}
\usepackage{amsthm}
\usepackage{caption}

\newtheorem{rema}{\sc Remark}[section]

\makeatletter      
\@addtoreset{equation}{section}
\makeatother       

\long \def\@makecation#1#2{ \vskip 10 pt
\setbox\@tempboxa\hbox{#1:#2} \ifdim \wd\@tempboxa >\hsize
\unhbox\@tempboxa\par \else \hbox to\hsize{\hfil\box\@temboxa\hfil}
\fi}

\def\hDash{\bot\!\!\!\bot}

\DeclareMathOperator*{\argmax}{arg\,max}

\begin{document}

\title{Transformed sufficient dimension reduction}

\author{Tao Wang$^1$, Xu Guo$^1$, Peirong Xu$^2$ and  Lixing Zhu$^1$\footnote{\scriptsize The corresponding author. Email: lzhu@hkbu.edu.hk. The research described here was supported by a grant from the Research Council of Hong Kong, and a grant from Hong Kong Baptist University, Hong Kong. The authors thank the Editor, the Associate editor and two referees for their constructive suggestions and comments that led to the significant improvement of an early manuscript. }\\ {\small \it $^1$Hong Kong Baptist University, Hong Kong, China}\\ {\small \it $^2$Southeast University, Nanjing, China}}
\date{}
\maketitle
\begin{abstract}
{A novel general framework is proposed in this paper for dimension reduction in regression to fill the gap between linear and fully nonlinear dimension reduction. The main idea is to transform first each of the raw predictors monotonically, and then search for a low-dimensional projection in the space defined by the transformed variables. Both user-specified and data-driven transformations are suggested. In each case, the methodology is discussed first in a general manner, and  a representative method, as an example, is then proposed and  evaluated by simulation. The proposed methods are applied to a real data set for illustration.}\\

\end{abstract}

\newpage

\section{Introduction}\label{sec1}

Consider the regression of a response variable $Y$ on a random vector $X = (X_1, \ldots, X_p)^T$ of predictors. In full generality the goal is to describe the dependence on $X$ of the conditional distribution of $Y$ given $X$. As remarked by Li (1991), lowering dimensionality prior to running a regression is practically important and in many cases crucial for further analysis: after projecting the data onto a smaller space, we are then in a better position to graphical displays, model building, curve fitting, model checking, and so on. For this purpose, linear sufficient dimension reduction (Cook 1998) focuses on finding a few linear combinations $\beta_1^T X, \ldots, \beta_d^T X$ that can replace $X$ without loss of information and without requiring a parametric model. If $\beta$ denotes the $p \times d$ matrix with columns $\beta_1, \ldots, \beta_d$, we require that
\begin{eqnarray*}
Y \hDash X|\beta^T X,
\end{eqnarray*}
where the notation $\hDash$ indicates independence. The statement is thus that $Y$ is independent of $X$ given any value assumed by $\beta^T X$, or equivalently, the conditional distribution of $Y|X$ equals that of $Y |\beta^T X$. If $\beta$ satisfies this relation, then its column space is called a dimension-reduction subspace (Li 1991). Under mild assumptions, the intersection of all dimension-reduction subspaces is itself a dimension-reduction subspace (Yin, Li and Cook 2008); in these cases it is called the central subspace for the regression of $Y$ on $X$, denoted by $S_{Y|X}$, and its dimension, $d_{Y|X} = {\rm dim}(S_{Y|X})$, is called the structural dimension. Another important and closely related concept is that of central mean subspace (Cook and Li 2002), which is concerned with dimension reduction for the conditional mean $E(Y|X)$. The conditional independence setting is
\begin{eqnarray*}
Y \hDash E(Y|X)|\beta^T X.
\end{eqnarray*}
The central mean subspace, written as $S_{E(Y|X)}$, is a proper subspace of the central subspace.

There are a variety of linear dimension-reduction methods in the literature. See, for example, Li (1991), Cook and Weisberg (1991), Li (1992), Cook and Li (2002), Xia et al. (2002), Cook and Ni (2005), Li and Wang (2007), Li, Wen and Zhu (2008), Li and Dong (2009), Zhu et al. (2010), Yin and Li (2011) and Ma and Zhu (2012). Among these methods, sliced inverse regression (SIR; Li 1991) and minimum average variance estimation (MAVE; Xia et al. 2002), which are proposed for estimating the central subspace and the central mean subspace respectively, are perhaps the most widely used. Another important issue is   structural dimension determination. Li (1991) provided a sequential test to help determine the number of significant SIR predictors; see also Bura and Cook (2001). Alternatively, one may use model selection criterion to consistently determine the dimension (Zhu, Miao and Peng 2006).

A more general paradigm of dimension reduction, termed nonlinear sufficient dimension reduction (Cook 2007), seeks an arbitrary function $\psi$ from $\mathbb{R}^p$ to $\mathbb{R}^d$ such that
\begin{eqnarray}
Y \hDash X | \psi(X) \quad (\mbox{or } \quad Y \hDash E(Y|X) | \psi(X)).
\end{eqnarray}
Several recent papers have proposed estimation procedures for nonlinear dimension reduction, which combine sufficient dimension reduction and machine learning techniques; see Wu (2008), Wu, Liang and Mukherjee (2008) and Li, Artemiou and Li (2011). The rationale is that if the data are concentrated on a nonlinear low-dimensional space, the linear dimension-reduction subspace estimated often has very large dimension. As a simple illustration, consider the regression model $Y = X_1 + X_2^2 + X_3^3 + \exp(X_4) + \epsilon$, where the dimension of $X$ is $p = 6$, and all the predictors and $\epsilon$ are independent. Linear sufficient dimension reduction provides $d_{Y|X} = 4$, whereas nonlinear sufficient dimension reduction gives $d = 1$ and $\psi(X) = X_1 + X_2^2 + X_3^3 + \exp(X_4)$, or any monotone function of it. Thus, in this example linear sufficient dimension reduction still suffers from the curse of dimensionality, because $d_{Y|X} = 4$ is pretty large although $d_{Y|X} < p$. However, it appears that by going from linear to nonlinear sufficient dimension reduction, the gain in generality is largely compensated by a loss of interpretability. {We note that  linear dimension-reduction methods are often used as the first step in statistical analysis: reducing dimensionality before undertaking another more sophisticated method. On the other hand,  nonlinear dimension-reduction methods may aim to solve the problem entirely at one stroke.} Consequently, it is desirable to set out a framework that offers a good compromise. This paper explores such a possibility.

Nonlinear transformation of variables is a commonly used practice in regression problems. For example, one is often tempted to use monotone transformation techniques to modify the response variable in a regression design (Box and Cox 1964). Concerning dimension reduction, however, transforming $X$ is always preferable to transforming $Y$, because the former just changes the way in which the conditional distribution of $Y|X$ is indexed. Cook (1998, Chapter 14) proposed graphical methods for visualizing predictor transformations that are useful for reducing the dimension of the central (mean) subspace. The development, however, was restricted to generalized linear models.

Given $p$ monotone univariate functions $f_1, \ldots, f_p$. Let $f(X) = (f_1(X_1), \ldots, f_p(X_p))^T$, or implicitly, $f = (f_1, \ldots, f_p)^T$. Then, (1.1) is equivalent to
\begin{eqnarray*}
Y \hDash f(X) | \varphi\{f(X)\} \quad (\mbox{or } \quad Y \hDash E\{Y|f(X)\} | \varphi\{f(X)\})
\end{eqnarray*}
for another function $\varphi$ from $\mathbb{R}^p$ to $\mathbb{R}^d$. To generalize linear sufficient dimension reduction while preserving its simplicity, we present a new framework by assuming that $\varphi$ is linear; that is, there exists an $p \times d$ matrix $B$ such  that
\begin{eqnarray}
Y \hDash f(X) | B^T f(X) \quad (\mbox{or } \quad Y \hDash E\{Y|f(X)\} | B^T f(X)).
\end{eqnarray}
We call this new paradigm transformed sufficient dimension reduction for the regression of $Y$ on $X$ with respect to $f$. Clearly, linear sufficient dimension reduction can be viewed as a special case, where $f_j = X_j$ for all $j = 1, \ldots, p$. Consider the illustrative example again. In terms of transformed sufficient dimension reduction, if we take $f_j= X_j$ for $j = 1, 2, 5$ and $6$, $f_3= X_3^3$ and $f_4 = \exp(X_4)$, then $d = 2$ and the dimensionality can not be further reduced. Further, since $f_j$ is monotone, $f_j(X_j)$ takes on the same general meaning as $X_j$ as an ``effect" predictor.

\section{Transformed sufficient dimension reduction}\label{sec3}

Before continuing, we should note that  each component function $f_j$ is unique only up to scale and shift. To ensure identifiability, we require that $\mu_f = E\{f(X)\} = 0$ and $\Sigma_f = {\rm{Cov}}\{f(X)\}$ is a correlation matrix whose main diagonal entries are equal to unity. We further assume, without loss of generality, that $f_j$ is a monotonically increasing function for any $j$.


A transformed dimension-reduction subspace for the regression of $Y$ on $X$ with respect to a given set of transformations $f = (f_1, \ldots, f_p)^T$ is any subspace $S \subseteq \mathbb{R}^p$ such that
$$Y \hDash f(X) | P_S f(X),$$
where $P_\cdot$ stands for the projection operator in the usual inner product. The intersection of all transformed dimension-reduction subspaces, provided itself satisfies this relation, is called the transformed central subspace for the regression of $Y$ on $X$ with respect to $f$, and is indicated with $S_{Y|f(X)}$. {Like $S_{Y|X}$, $S_{Y|f(X)}$ uniquely exists under very mild conditions, and is assumed to exist throughout this paper.} Its dimension, say, $d_{Y|f(X)} = {\rm{dim}}(S_{Y|f(X)})$, is still called the structural dimension.

When only the mean response is of interest, transformed sufficient dimension reduction can be defined in a similar fashion. A transformed mean dimension-reduction subspace for the regression of $Y$ on $X$ with respect to $f = (f_1, \ldots, f_p)^T$ is any subspace $S \subseteq \mathbb{R}^p$ such that
$$Y \hDash E\{Y|f(X)\} | P_S f(X).$$
If the intersection of all transformed mean dimension-reduction subspaces is also a transformed mean dimension-reduction subspace, it is called the transformed central mean subspace for the regression of $Y$ on $X$ with respect to $f$, and is written as $S_{E\{Y|f(X)\}}$.

\begin{rema}
Throughout the paper, we tacitly assume that the predictors are continuous and  the transformation functions are smooth. Further, we exclude from the analysis singular functions such as the Cantor function.
\end{rema}

\begin{rema}
Transformed sufficient dimension reduction, which lies between linear and nonlinear sufficient dimension reduction, has the ease of interpretation of the former and retains the flexibility of the latter. The idea is to apply linear dimension-reduction methods in the population after replacing $X$ by $f(X)$, and in the sample once the $f_j$'s have been specified and/or estimated. However, transformed sufficient dimension reduction inherits one drawback of nonlinear sufficient dimension reduction that the transformation needs not be unique. To see this, consider again the regression model $Y = X_1 + X_2^2 + X_3^3 + \exp(X_4) + \epsilon$. In terms of transformed sufficient dimension reduction, $d_{Y|f(X)} = 2$ and, up to scale and shift, $f_1 = X_1, f_3 = X_3^3$ and $f_4 = \exp(X_4)$, but for $j = 2, 5$ and $6$, $f_j$ can be any monotone function of $X_j$. Nevertheless, the non-identifiability of transformations of this type is not a fatal flaw, because the transformed central (mean) subspace is still well-defined, and thus all of the infinite many  transformations are feasible. Because we are not assuming a model for $Y|X$, this needs not change the fundamental issues in regression. More precisely, since $Y|X$ has the same distribution as $Y|f(X)$, different transformations just change the way in which the conditional distribution of $Y|X$ is indexed.
\end{rema}

\section{Estimation: user-specified transformations}\label{sec33}

If the distribution of $X$ is not normal, it is sometimes convenient to consider transformations that help to normalize the observed data; for example, by taking a certain power, or  the logarithm. Two well-known and widely-used parametric transformations are the Box-Cox transformation (Box and Cox 1964) and the Yeo-Johnson transformation (Yeo and Johnson 2000). In this section we assume that the transformed vector $f(X) = (f_1(X_1), \ldots, f_p(X_p))^T$ is multivariate Gaussian. Under the identifiability condition, $f_j = \Phi^{-1}(F_j^X)$, where $F_j^X$ and $\Phi$ denote, respectively, the marginal distribution function of $X_j$ and the one-dimensional standard normal distribution function. The transformations used here are usually referred to as the probability integral transformations and form a standard tool in simulation methodology. Since the $f_j$'s have been specified, one can proceed by invoking the linear dimension-reduction methods described in the introduction. Below we focus on SIR.

\subsection{Probability-integral-transformed sliced inverse regression}\label{sec332}

Assume that the data $(x_i, y_i), i = 1, \ldots, n$, are independent and identically distributed observations on $(X, Y)$, where $x_i = (x_{i1}, \ldots, x_{ip})^T$. Let $\hat{F}_{j}^X$ be an estimator of $F_j^X$. Define the normal scores $\hat{f}_j(x_{ij}) = \Phi^{-1}\{\hat{F}_j^X(x_{ij})\}$ for $i = 1, \ldots, n$ and $j = 1, \ldots, p$. Probability-integral-transformed SIR uses a two-step procedure:
\begin{enumerate}
\item[S1.] Replace the observations, for each predictor, by their corresponding normal scores.
\item[S2.] Apply SIR to the transformed data to estimate $S_{Y|f(X)}$ and ascertain its dimension.
\end{enumerate}
Currently, the most popular estimator of $F_j^X$ is the empirical distribution function. In order to avoid difficulties arising from the potential unboundedness of $\Phi^{-1}(t)$ as $t$ tends to one, we adopt instead the rescaled empirical distribution function
$$\hat{F}_j^X(t) = \frac{1}{1 + n}\sum_{i = 1}^n I(x_{ij} \leq t),$$
where $I(\cdot)$ is the indicator function. 

In the following we give the implementation details of the second step. We work in the scale of the standardized predictor $Z^f = \Sigma_f^{-1/2} f(X)$, because $S_{Y|f(X)} = \Sigma_f^{-1/2}S_{Y|Z^f}$ (Cook 1998, Proposition 10.1). Since $f(X)$ is multivariate Gaussian, the matrix ${\rm{Cov}}\{E(Z^f|Y)\}$ is degenerate in any direction orthogonal to $S_{Y|Z^f}$. For simplicity, we assume that $S_{Y|Z^f}$ coincides with the column space of ${\rm{Cov}}\{E(Z^f|Y)\}$.

Let
$$\bar{f} = \frac{1}{n}\sum_{i = 1}^n\hat{f}(x_i) \quad {\rm{and}} \quad \hat{\Sigma}_{f} = \frac{1}{n}\sum_{i = 1}^n\{\hat{f}(x_i) - \bar{f}\}\{\hat{f}(x_i) - \bar{f}\}^T,$$
where $\hat{f}(x_i) = (\hat{f}_1(x_{i1}), \ldots, \hat{f}_p(x_{ip}))^T$. Define $\hat{z}_i^f = \hat{\Sigma}_f^{-1 / 2}\{\hat{f}(x_i) - \bar{f}\}$. We divide the range of $Y$ into $H$ slices, and calculate the sample mean of the $\hat{z}_i^f$'s within each slice as
$$\bar{z}_h^f = \frac{1}{n_h}\sum_{i|h} \hat{z}_i^f, \quad h = 1, \ldots, H,$$
where the summation is over the indices $i$ of the $y_i$'s that fall into slice $h$, and $n_h$ is the number of observations in that slice. Probability-integral-transformed SIR estimates ${\rm{Cov}}\{E(Z^f|Y)\}$ by
$$\widehat{\rm{Cov}}\{E(Z^f|Y)\} =
\frac{1}{n}\sum_{h = 1}^H n_h \bar{z}_h^f \bar{z}_h^{fT}.$$

Let $\hat{\lambda}_1^f \geq \cdots \geq \hat{\lambda}_p^f$ be the ordered eigenvalues of $\widehat{\rm{Cov}}\{E(Z^f|Y)\}$ and let $\hat{v}_1^f, \ldots, \hat{v}^f_{d_{Y|f(X)}}$ be the eigenvectors corresponding to the $d_{Y|f(X)}$ largest eigenvalues. The estimators of the directions in $S_{Y|f(X)}$ are $\hat{\eta}_j^f = \hat{\Sigma}_f^{-1/2}\hat{v}_j^f$, and the probability-integral-transformed SIR predictors are given by $\hat{\eta}_j^{fT} f(X), j = 1, \ldots, d_{Y|f(X)}$. To infer about the structural dimension $d_{Y|f(X)}$, we use the method suggested by Li (1991) and Bura and Cook (2001). Specifically, we use the test statistic
$$L_{d}^f = n \sum_{j = d + 1}^p \hat{\lambda}_j^f.$$
Starting with $d = 0$, test the hypothesis $d_{Y|f(X)} = d$ versus $d_{Y|f(X)} > d$. If the test is rejected, increment $d$ by one and test again, stopping with the first nonsignificant result. Although they are practically useful, sequential tests generally yield a decision of the structural dimension that depends on the nominal significance levels, and thus are not consistent (Zhu, Miao and Peng 2006). It is nevertheless possible to estimate the structural dimension directly. To this end, we adopt the modified BIC-type criterion of Zhu et al. (2010) that is based on Zhu, Miao and Peng (2006). Define
$${\rm{BIC}}_{d}^f = \frac{\sum_{j = 1}^d \hat{\lambda}_{j}^{f2}}{\sum_{j = 1}^p \hat{\lambda}_l^{f2}} - \frac{\kappa_n}{n} \times \frac{d(d + 1)}{2},$$
where $\kappa_n$ is a penalty factor, and $d(d + 1)/2$ denotes the number of free parameters when the matrix ${\rm{Cov}}\{E(Z^f|Y)\}$ is of rank $d$. The estimated structural dimension is then
$$\hat{d}_{Y|f(X)} = \argmax_{1 \leq d \leq p} {\rm{BIC}}_d^f.$$

Computationally, probability-integral-transformed SIR is not more difficult than SIR, because one can exploit existing software for SIR and the only additional cost is the estimation of the transformations. Fortunately for us, the first step is non-iterative, and has the advantage of making fewer assumptions and being easier to compute than parametric transformations. Theoretical properties of probability-integral-transformed SIR, assuming that the transformations are correctly specified, are provided in the supplementary material.

\begin{rema}
Response transformations, such as the slicing technique in the SIR algorithm, are exploited to suggest interesting patterns in the data. However, a disadvantage of SIR is that, unlike additive regression models, transformations are  not allowed to make for all the predictors separately (Chen and Li 1998, p. 296). To this end, nonlinear multivariate techniques, such as ACE of Breiman and Friedman (1985), allow transformations on both the response variable and the predictors. Only one transformation on $Y$, however, is allowed in the ACE algorithm. In this regard, probability-integral-transformed SIR provides a certain remedy.
\end{rema}

\begin{rema}\label{robustrema}
Using user-specified transformations directly gives rise to simple and fast estimation procedures, and our numerical results in the next subsection show that probability integral transformations are more flexible than the parametric ones. However, this approach is not entirely free of the problem of misspecification. In particular, the assumption regarding the distribution of $f(X)$ can be easily violated in many applications. In the next subsection, we also illustrate the robustness of probability-integral-transformed SIR against nonnormality by simulation. To address the problem of  misspecification, the transformations have to be estimated fully nonparametrically, and a general framework is developed in Section \ref{sec4}. Unfortunately, these robust procedures are iterative and computationally demanding.
\end{rema}

\subsection{Simuilation results}

In this section we use a simulation study to investigate the performance of probability-integral-transformed SIR. Consider the following model
\begin{eqnarray}\label{modelsir}
Y = \frac{f_1 + f_2}{(f_3 + f_4 + 1.5)^2 + 0.5} + 0.5 \epsilon,
\end{eqnarray}
where $f = (f_1, \ldots, f_p)^T \sim N(0, \Sigma_f)$ with $(\Sigma_f)_{ij}
= 0.5^{|i - j|}$ for $ 0 \leq i, j \leq p = 10$, $\epsilon \sim N(0, 1)$, and $f$ and $\epsilon$ are independent. 

Six different cases are explored to sample data from transformed Gaussian distributions. We first generate $f = f(X)$ from $N(0, \Sigma_f)$, and then use either power transformation (Case 1) or probability integral transformation $X_j = F_j^{X-1}\{\Phi(f_j)\}$ (Cases 2-6) to generate $X$. The details are as follows. 

\noindent {\sc Case 1.} $X_j = {\rm{sign}}(f_j) \times f_j^2$ for $j = 1, \ldots, 10$.\\
\noindent {\sc Case 2.} $X_j$ has a central skew-Laplace distribution with parameters 2 and 6 for all $j$.\\
\noindent {\sc Case 3.} $X_j$ has a beta distribution with parameters 3 and 0.5 for $j = 1, \ldots, 3$ and $X_j$ has an exponential distribution with mean 1 for $j = 4, \ldots, 10$.\\
\noindent {\sc Case 4.} $X_j$ has a $t$-distribution with $k$ degrees of freedom: $k = 2$ for $j = 1, \ldots, 3$, $k = 3$ for $j = 4, \ldots, 6$ and $k = 4$ for $j = 7, \ldots, 10$.\\
\noindent {\sc Case 5.} $X_j$ has a normal mixture distribution with the outlier density, $\#5$, used in Marron and Wand (1992) for all $j$.\\
\noindent {\sc Case 6.} $X_j$ has a standard Cauchy distribution for all $j$.

In each case, we generate 200 datasets with the sample size $n = 200$ and $n = 400$. For the regression of $Y$ on $f(X)$, the structural dimension is $d_{Y|f(X)} = 2$; we evaluate the performance of SIR assuming that $f$ is known, probability-integral-transformed SIR, and SIR after the Yeo-Johnson transformation. We use the Yeo-Johnson transformation because it is well-defined on the whole real line and has properties similar to those of the Box-Cox transformation. For comparison, we also examine SIR for the regression of $Y$ on $X$; in this case the structural dimension is $d_{Y|X} = 4$. The resulting estimators are denoted respectively by f-SIR, T-SIR, YJ-SIR and SIR. Ten slices are used for all four methods considered here; it is well-known that the performance of SIR is not very sensitive to the number of slices, although how to tune the number of slices remains a difficult open problem.

Both the vector correlation coefficient (VCC) and the trace correlation coefficient (TCC) are employed to evaluate the estimation accuracy. For an estimator $\hat{B}$ of $B$, VCC is defined to be $(\prod_{l = 1}^d \phi_l^2)^{1 / 2}$ and TCC is defined to be $(d^{-1}\sum_{l = 1}^d \phi_l^2)^{1 / 2}$, where $1 \geq \phi_1^2 \geq \cdots \geq \phi_d^2 \geq 0$ are the eigenvalues of the matrix $\hat{B}_o^T B_o B_o^T \hat{B}_o$ with $\hat{B}_o$ and $B_o$ being the orthonormalized versions of $\hat{B}$ and $B$ respectively. A correlation coefficient closer to unity means better estimation of the (transformed) central subspace. Here, for $Y|f(X)$ we have $B = (\eta_1^f, \eta_2^f)$ and $d = 2$, while for $Y|X$ we have $B = (e_1, e_2, e_3, e_4)$ and $d = 4$, where $\eta_1^f = (1, 1, 0, \ldots, 0)^T, \eta_2^f = (0, 0, 1, 1, 0, \ldots, 0)^T$ and $e_i$ is a vector of length 10 whose $i$-th element is 1 and all other elements are 0. The means and standard deviations of VCC and TCC, based on 200 repetitions, are presented in Tables \ref{tab11} and \ref{tab12}. Several observations can be made as follows. First, we can see that, somewhat surprisingly, T-SIR performs slightly better than f-SIR; that is, using the estimated transformations yields a more accurate estimate than using the true ones. Second, we observe that, except for Case 3, probability-integral-transformed SIR is the best performer, followed by f-SIR and YJ-SIR. As we can see, the performance of SIR after the Yeo-Johnson transformation is very sensitive to the marginal distributions of $X$; YJ-SIR performs poorly in Cases 5 and 6. Third, using SIR directly for the regression of $Y$ on $X$ leads to very poor estimates with alarmingly low vector correlation coefficient. This is not unexpected, because $d_{Y|X} = 4$ is much larger than $d_{Y|f(X)} = 2$, making the estimation problem considerably more difficult. Finally, as the sample size increases, the performance of f-SIR, T-SIR and YJ-SIR improves greatly, but that of SIR is not much affected.

\begin{table}[htb!]\caption{The means and standard deviations (in parentheses) of the vector correlation coefficient (VCC) and the trace correlation coefficient (TCC), based on 200 repetitions, are reported for various estimators when $n = 200$} \label{tab11} \vspace{-0.3cm}
\centering
 {\small\scriptsize\hspace{12.5cm}
\renewcommand{\arraystretch}{1} \tabcolsep 0.15cm
\begin{tabular}{ccccc}
\hline &  \multicolumn{1}{c}{VCC}  &  \multicolumn{1}{c}{TCC}  & \multicolumn{1}{c}{VCC}  &  \multicolumn{1}{c}{TCC}\\
       & \multicolumn{2}{c}{f-SIR} & \multicolumn{2}{c}{T-SIR}\\
       & 0.6537 (0.1662) & 0.8354 (0.0642) & 0.6695 (0.1539) &  0.8416 (0.0614)\\
       & \multicolumn{2}{c}{YJ-SIR} & \multicolumn{2}{c}{SIR}\\
Case 1 & 0.4301 (0.2086) &  0.7447 (0.0755) & 0.1132 (0.1029) & 0.7321 (0.0535)\\
Case 2 & 0.5887 (0.1905) &  0.8083 (0.0724) & 0.1336 (0.1222) & 0.7576 (0.0495)\\
Case 3 & 0.7114 (0.1361) &  0.8598 (0.0556) & 0.4285 (0.2454) & 0.8807 (0.0431)\\
Case 4 & 0.4570 (0.2012) &  0.7476 (0.0783) & 0.0463 (0.0538) & 0.6653 (0.0601)\\
Case 5 & 0.3242 (0.2006) &  0.6998 (0.0747) & 0.0981 (0.0882) & 0.7169 (0.0542)\\
Case 6 & 0.2322 (0.1728) &  0.6292 (0.0930) & 0.0446 (0.0926) & 0.6656 (0.0941)\\
\hline
\end{tabular} }
\end{table}

\begin{table}[htb!]\caption{The means and standard deviations (in parentheses) of the vector correlation coefficient (VCC) and the trace correlation coefficient (TCC), based on 200 repetitions, are reported for various estimators when $n = 400$} \label{tab12} \vspace{-0.3cm}
\centering
 {\small\scriptsize\hspace{12.5cm}
\renewcommand{\arraystretch}{1} \tabcolsep 0.15cm
\begin{tabular}{ccccc}
\hline &  \multicolumn{1}{c}{VCC}  &  \multicolumn{1}{c}{TCC}  & \multicolumn{1}{c}{VCC}  &  \multicolumn{1}{c}{TCC}\\
       & \multicolumn{2}{c}{f-SIR} & \multicolumn{2}{c}{T-SIR}\\
       & 0.8178 (0.0920) & 0.9106 (0.0416) & 0.8230 (0.0885) &  0.9130 (0.0405)\\
       & \multicolumn{2}{c}{YJ-SIR} & \multicolumn{2}{c}{SIR}\\
Case 1 & 0.6348 (0.1705) &  0.8316 (0.0646) & 0.1472 (0.1186) & 0.7610 (0.0469)\\
Case 2 & 0.7778 (0.1137) &  0.8921 (0.0487) & 0.1710 (0.1468) & 0.7837 (0.0462)\\
Case 3 & 0.8483 (0.0653) &  0.9245 (0.0313) & 0.4648 (0.2511) & 0.8894 (0.0428)\\
Case 4 & 0.6363 (0.1569) &  0.8255 (0.0654) & 0.0519 (0.0568) & 0.6907 (0.0524)\\
Case 5 & 0.4543 (0.2215) &  0.7661 (0.0711) & 0.1038 (0.0999) & 0.7353 (0.0531)\\
Case 6 & 0.2695 (0.1867) &  0.6628 (0.0846) & 0.0474 (0.0895) & 0.6780 (0.1070)\\
\hline
\end{tabular} }
\end{table}

The estimation accuracy of each method relies on the selection of the structural dimension which is assumed to be known in the above discussion. We now study numerical aspects of the two aforementioned methods for determining the dimension: the sequential test with nominal significance level 0.05 and the BIC-type criterion with $\kappa_n = \log(n)$. The empirical counts out of 200 repetitions are reported in Tables \ref{tab13} and \ref{tab14}. As we can see, SIR for $Y|X$ tends to consistently underestimate the structural dimension. For f-SIR, T-SIR and YJ-SIR, the BIC-type criterion outperforms the sequential test, indicating that the test procedure is not consistent. Regarding the BIC-type criterion, generally, f-SIR and T-SIR perform comparably well, followed by YJ-SIR whose performance suffers in Cases 5 and 6.

\begin{table}[htb!]\caption{The frequencies of estimated structural dimension out of 200 repetitions by the sequential test (TEST) and the BIC-type criterion when $n = 200$} \label{tab13} \vspace{-0.3cm}
\centering
 {\small\scriptsize\hspace{12.5cm}
\renewcommand{\arraystretch}{1} \tabcolsep 0.1cm
\begin{tabular}{cccccccc}
\hline&      & \multicolumn{3}{c}{TEST} & \multicolumn{3}{c}{BIC}\\
      &      & $\hat{d}_{Y|f(X)} < 2$ & $\hat{d}_{Y|f(X)} = 2$ & $\hat{d}_{Y|f(X)} > 2$ & $\hat{d}_{Y|f(X)} < 2$ & $\hat{d}_{Y|f(X)} = 2$ & $\hat{d}_{Y|f(X)} > 2$ \\
f-SIR &      &  88& 107 & 5  & 1 & 197 &   2\\
T-SIR &      &  83& 111 & 6  & 3 & 196 &   1\\
YJ-SIR&Case~1& 153&  45 & 2  &  5& 183 &  12\\
      &Case~2& 109&  88 & 3  &  1& 196 &   3\\
      &Case~3&  76& 119 & 5  &  4& 196 &   0\\
      &Case~4& 142&  54 & 4  &  3& 172 &  25\\
      &Case~5& 178&  20 & 2  &  1& 140 &  59\\
      &Case~6& 188&  12 & 0  &  0&  77 & 123\\
      &      & $\hat{d}_{Y|X} < 4$ & $\hat{d}_{Y|X} = 4$ & $\hat{d}_{Y|X} > 4$ & $\hat{d}_{Y|X} < 4$ & $\hat{d}_{Y|X} = 4$ & $\hat{d}_{Y|X} > 4$ \\
SIR   &Case~1& 200&   0 & 0  &200&   0 &   0\\
      &Case~2& 200&   0 & 0  &200&   0 &   0\\
      &Case~3& 200&   0 & 0  &200&   0 &   0\\
      &Case~4& 200&   0 & 0  &199&   1 &   0\\
      &Case~5& 200&   0 & 0  &199&   1 &   0\\
      &Case~6& 200&   0 & 0  &188&  12 &   0\\
\hline
\end{tabular} }
\end{table}

\begin{table}[htb!]\caption{The frequencies of estimated structural dimension out of 200 repetitions by the sequential test (TEST) and the BIC-type criterion when $n = 400$} \label{tab14} \vspace{-0.3cm}
\centering
 {\small\scriptsize\hspace{12.5cm}
\renewcommand{\arraystretch}{1} \tabcolsep 0.1cm
\begin{tabular}{cccccccc}
\hline&      & \multicolumn{3}{c}{TEST} & \multicolumn{3}{c}{BIC}\\
      &      & $\hat{d}_{Y|f(X)} < 2$ & $\hat{d}_{Y|f(X)} = 2$ & $\hat{d}_{Y|f(X)} > 2$ & $\hat{d}_{Y|f(X)} < 2$ & $\hat{d}_{Y|f(X)} = 2$ & $\hat{d}_{Y|f(X)} > 2$ \\
f-SIR &      &   6& 192 & 2  & 0 & 200 &   0\\
T-SIR &      &   6& 192 & 2  & 0 & 200 &   0\\
YJ-SIR&Case~1&  88& 108 & 4  &  6& 189 &   5\\
      &Case~2&  20& 177 & 3  &  1& 199 &   0\\
      &Case~3&   3& 195 & 2  &  2& 198 &   0\\
      &Case~4&  72& 124 & 4  &  0& 186 &  14\\
      &Case~5& 149&  49 & 2  &  3& 166 &  31\\
      &Case~6& 188&  12 & 0  &  0&  38 & 162\\
      &      & $\hat{d}_{Y|X} < 4$ & $\hat{d}_{Y|X} = 4$ & $\hat{d}_{Y|X} > 4$ & $\hat{d}_{Y|X} < 4$ & $\hat{d}_{Y|X} = 4$ & $\hat{d}_{Y|X} > 4$ \\
SIR   &Case~1& 200&   0 & 0  &200&   0 &   0\\
      &Case~2& 199&   1 & 0  &200&   0 &   0\\
      &Case~3& 200&   0 & 0  &200&   0 &   0\\
      &Case~4& 200&   0 & 0  &200&   0 &   0\\
      &Case~5& 200&   0 & 0  &200&   0 &   0\\
      &Case~6& 200&   0 & 0  &151&  49 &   0\\
\hline
\end{tabular} }
\end{table}

We now carry out a sensitivity analysis to evaluate the performance of probability-integral-transformed SIR when the normality assumption is not met. We consider again the model in (\ref{modelsir}), expect that the distribution of $f$ is non-Gaussian. Specifically, we concentrate on the following two cases.

\noindent {\sc Case 7.} $f \sim t_k(0, \Sigma_f)$. That is, $f$ has a $t$-distribution with location vector 0, scale matrix $\Sigma_f$ and $k$ degrees of freedom. Three values of $k$ are explored: 5, 10 and 20. It is well known that for $k \rightarrow \infty$ the $t$-distribution approaches a normal distribution, but for $k < \infty$ the $t$-distribution has ``fatter tails" than the corresponding normal distribution. \\
\noindent {\sc Case 8.} $f = \Sigma_f^{1/2} u$, and $u = (u_1, \ldots, u_{10})^T$ is uniform on a 10-dimensional cube $[-\sqrt{3}, \sqrt{3}]^{10}$. The normality assumption is seriously violated in this case.

The results, calculated from 200 simulated samples, are summarized in Tables \ref{robust1} and \ref{robust2}. We see that probability-integral-transformed SIR achieves a degree of robustness against non-Gaussianity of the distribution of $f$.

\begin{table}[htb!]\caption{The means and standard deviations (in parentheses) of the vector correlation coefficient (VCC) and the trace correlation coefficient (TCC), based on 200 repetitions, are reported} \label{robust1} \vspace{-0.3cm}
\centering
 {\small\scriptsize\hspace{12.5cm}
\renewcommand{\arraystretch}{1} \tabcolsep 0.15cm
\begin{tabular}{cccccc}
\hline & & \multicolumn{2}{c}{$n = 200$} & \multicolumn{2}{c}{$n = 400$}\\
       & & \multicolumn{1}{c}{VCC} & \multicolumn{1}{c}{TCC} & \multicolumn{1}{c}{VCC} & \multicolumn{1}{c}{TCC}\\
T-SIR  &Case 7 ($k = 5$) &    0.6103 (0.1550) & 0.8155 (0.0604) & 0.7867 (0.0880) & 0.8953 (0.0405)\\
       &Case 7 ($k = 10$)&    0.6312 (0.1740) & 0.8248 (0.0705) & 0.8101 (0.0860) & 0.9069 (0.0392)\\
       &Case 7 ($k = 20$)&    0.6611 (0.1459) & 0.8364 (0.0613) & 0.8171 (0.0870) & 0.9102 (0.0406)\\
       &Case 8           &    0.6546 (0.1570) & 0.8340 (0.0632) & 0.8127 (0.0768) & 0.9079 (0.0356)\\
f-SIR  &Case 7 ($k = 5$) &    0.5129 (0.1832) & 0.7739 (0.0677) & 0.7158 (0.1139) & 0.8624 (0.0496)\\
       &Case 7 ($k = 10$)&    0.5918 (0.1761) & 0.8064 (0.0718) & 0.7790 (0.1002) & 0.8925 (0.0447)\\
       &Case 7 ($k = 20$)&    0.6404 (0.1527) & 0.8267 (0.0650) & 0.7984 (0.0979) & 0.9016 (0.0444)\\
       &Case 8           &    0.7173 (0.1401) & 0.8629 (0.0568) & 0.8533 (0.0614) & 0.9274 (0.0291)\\
\hline
\end{tabular} }
\end{table}

\begin{table}[htb!]\caption{The frequencies of estimated structural dimension out of 200 repetitions by the BIC-type criterion} \label{robust2} \vspace{-0.3cm}
\centering
 {\small\scriptsize\hspace{12.5cm}
\renewcommand{\arraystretch}{1} \tabcolsep 0.1cm
\begin{tabular}{cccccccc}
\hline&      & \multicolumn{3}{c}{$n = 200$} & \multicolumn{3}{c}{$n = 400$}\\
      &      & $\hat{d}_{Y|f(X)} < 2$ & $\hat{d}_{Y|f(X)} = 2$ & $\hat{d}_{Y|f(X)} > 2$ & $\hat{d}_{Y|f(X)} < 2$ & $\hat{d}_{Y|f(X)} = 2$ & $\hat{d}_{Y|f(X)} > 2$ \\
T-SIR  &Case 7 ($k = 5$) &   2& 192 &  6 &  0& 199 &   1\\
       &Case 7 ($k = 10$)&   4& 192 &  4 &  1& 199 &   0\\
       &Case 7 ($k = 20$)&   0& 197 &  3 &  0& 200 &   0\\
       &Case 8           &   2& 195 &  3 &  3& 200 &   0\\
f-SIR  &Case 7 ($k = 5$) &   4& 186 & 10 &  1& 198 &   1\\
       &Case 7 ($k = 10$)&   6& 190 &  4 &  2& 197 &   1\\
       &Case 7 ($k = 20$)&   1& 196 &  3 &  0& 200 &   0\\
       &Case 8           &   1& 197 &  2 &  0& 200 &   0\\
\hline
\end{tabular} }
\end{table}

\section{Estimation: data-driven transformations}\label{sec4}

As mentioned in Remark \ref{robustrema}, the above estimation procedure, which makes use of the well-known probability integral transformation, suffers from misspecification of transformations. One strategy for dealing with this problem is to estimate the monotone transformations fully nonparametrically, as shown below.

Since a monotonically increasing function has a positive first derivative, it is reasonable to express $f_j$ as
\begin{eqnarray}\label{monorep}
f_j(t) = C_j + \int_{t_{j1}}^{t} \exp\{s_j(t)\} dt,
\end{eqnarray}
where $C_j$ is a constant, $s_j$ is an unconstrained function, and $t_{j1}$ is the fixed origin for the range of $t$-values for which the data are being fit. For simplicity, we assume that for each $j$, $s_j(t)$ can be well approximated by a linear combination of basis functions $\{1, \theta_{j1}(t), \ldots, \theta_{jM}(t)\}$,
$$s_j(t) = c_{j0} + \sum_{m = 1}^{M}c_{jm}\theta_{jm}(t), \quad j = 1, \ldots, p.$$
In this regard, the principal advantage brought by representation (\ref{monorep}) is the conversion of the estimation problem from a problem of finding the constrained function $f_j$ to a problem of computing the unconstrained function $s_j$. For details regarding monotone transformation techniques and their applications, see Ramsay (1988) and Ramsay and Silverman (2005).

Although other methods can be adapted along the lines developed here, we concentrate on MAVE. In particular, we consider the transformed sufficient dimension-reduction problem motivated by the model
\begin{equation}\label{mavemodel}
Y = Q\{B^T f(X)\} + \epsilon,
\end{equation}
where $Q$ is an unknown smooth link function, $B$ is a $p \times d$ orthogonal matrix with $d$ being the structural dimension, and $E(\epsilon|X) = 0$ almost surely. We first assume that $d$ is known.

\subsection{Monotone-smoothing-transformed minimum average variance estimation}\label{sec42}

Xia et al. (2002) proposed MAVE for a special case of model (\ref{mavemodel}), where $f_j(t) = t$ for all $j$. The MAVE method has been found very useful in semi-parametric estimation and linear dimension reduction. It is easy to implement and can be easily adapted in various ways to suit special statistical requirements, such as robust regression, feature selection and censored data. We next discuss its application to nonlinear dimension reduction.


The new procedure combines MAVE and monotone spline smoothing with a roughness penalty. For simplicity, we assume that the argument values for the $k$-th transformation function $f_k$ are within the interval $[t_{k1}, t_{k2}]$. 
Let $f^{ij} = f(x_i) - f(x_j)$ and let $f^{ij}_{k}$ denote the $k$-th element of $f^{ij}$. The fitting criterion considered here is
\begin{eqnarray}\label{4.2.1}
\frac{1}{n}\sum_{j = 1}^n\sum_{i = 1}^n (y_i - a_j - b_j^T
B^T f^{ij})^2 w_{ij} + \lambda \sum_{k = 1}^p \int_{t_{k1}}^{t_{k2}} \{D^2s_k(t)\}^2 dt,
\end{eqnarray}
where $\lambda$ is a smoothing parameter, and we use the notation $D$ for differentiation.

For convenience we let $c_k = (c_{k0}, c_{k1}, \ldots, c_{km})^T$ and $\Theta_k(t) = (1, \theta_{k1}(t), \ldots, \theta_{kM}(t))^T$. We let $a = (a_1, \ldots, a_n)^T, b = (b_1^T, \ldots, b_n^T)^T$ and $c = (c_1^T, \ldots, c_p^T)^T.$ Write $B = (B_1, \ldots, B_p)^T$ with $B_k$ denoting the $k$-th row of $B$. We standardize each predictor to have zero mean and unit variance. The optimization for (\ref{4.2.1}) can be stated as follows.

Beginning with an initial estimate $c^{(0)}$, which may be a vector of zeros, estimate $a^{(0)}, b^{(0)}$ and $B^{(0)}$ by the MAVE procedure. On any iteration $v > 0$, let $a^{(v - 1)}, b^{(v - 1)}$ and $B^{(v - 1)}$ be the estimates from the previous iteration. Set $v = 1$, we proceed as follows.

\begin{itemize}\setlength{\itemindent}{1.1em}
\item[Step~1.] Fixing $a = a^{(v - 1)}, b = b^{(v - 1)}$ and $B = B^{(v - 1)}$, use the $p$-block Gauss-Seidel scheme to calculate $c^{(v)}$. Set $\tau = 0$ and $c_{k}^{(v - 1, \tau)} = c_{k}^{(v - 1)}$.
    \begin{itemize}\setlength{\itemindent}{1.0em}
    \item[Step~1.1.] For $l = 1, \ldots, p$, let $c_{-l} = (c_1^T, \ldots, c_{l - 1}^T, c_{l + 1}^T, \ldots, c_p^T)^T$ and write (\ref{4.2.1}) as
        \begin{multline*}
        \Gamma(c_l; c_{-l}) = \frac{1}{n}\sum_{j = 1}^n\sum_{i = 1}^n \left\{y_i - a_j - \sum_{k \neq l} b_j^T B_k f^{ij}_{k} - b_j^T B_l f^{ij}_{l} \right\}^2 w_{ij}\\ + \lambda \sum_{k \neq l}^p \int_{t_{k1}}^{t_{k2}} \{D^2s_k(t)\}^2 dt + \lambda \int_{t_{l1}}^{t_{l2}} \{D^2s_l(t)\}^2 dt.
        \end{multline*}
        Fixing $c_{-l} = (c_1^{(v - 1, \tau + 1)}, \ldots, c_{l - 1}^{(v - 1, \tau + 1)}, c_{l + 1}^{(v - 1, \tau)}, \ldots, c_p^{(v - 1, \tau)})$, optimize $\Gamma(c_l; c_{-l})$ with respect to $c_l$ by the Gauss-Jordan or scoring procedure for non-linear least squares problems to obtain $c_l^{(v - 1, \tau + 1)}$. Specifically, the Gauss-Jordan procedure (see, e.g., Ramsay 1998) requires that the update vector
        $$\delta^{(u + 1)} = c_l^{(u + 1)} - c_l^{(u)}$$
        be the solution of the linear equation
        $$H_l^{(u)} \delta^{(u + 1)} = - s_l^{(u)},$$
        where
        $$H_l = \frac{1}{n}X_l^{*T} X_l^{*} + \lambda P_l, \ s_l = -\frac{1}{n} X_l^{*T}r^* + \lambda P_l c_l,$$
        matrix $X_l^*$ is $n^2 \times m$ and has rows $$(b_j^T B_l) \sqrt{w_{ij}} \int_{t_{l1}}^{x_{il} - x_{jl}} \Theta_l(s) \exp\{c_l ^T \Theta_l(s)\} ds,$$
        matrix $P_l$ of order $m \times m$ is $$\int_{t_{l1}}^{t_{l2}} D^2\Theta_l(s) \{D^2\Theta_l(s)\}^T ds,$$
        and $r^*$ is the residual vector of length $n^2$ with elements $(y_i - a_j - b_j^T B^T f^{ij}) \sqrt{w_{ij}}$.
    \item[Step~1.2.] If a convergence criterion is met, stop and set $c^{(v)} = (c_1^{(v - 1, \tau + 1)}, \ldots, c_{p}^{(v - 1, \tau + 1)})$; otherwise, set $\tau$ to be $\tau + 1$ and go to Step~1.1.
    \end{itemize}
\item[Step~2.] Fixing $a = a^{(v - 1)}, b = b^{(v - 1)}$ and $c = c^{(v)}$, calculate the solution of $B$ to (\ref{4.2.1}):
    \begin{eqnarray*}
    {\vec{B}}^{(v)} = \left\{\sum_{j = 1}^{n}\sum_{i = 1}^{n} w_{ij} (f^{ij} \otimes b_j) (f^{ij} \otimes b_j)^T\right\}^{-1} \times \sum_{j = 1}^{n}\sum_{i = 1}^{n} w_{ij} (y_i - a_j),
    \end{eqnarray*}
    where $\vec{B} = {\rm vec}(B^T)$ with ${\rm vec}(\cdot)$ being a matrix operator that stacks all columns of a matrix into a vector. Standardize each monotonically transformed predictor to have zero mean and unit variance, and normalize $B^{(v)}$ such that $B^{(v)}B^{(v)T} = I_d$.
\item[Step~3.] Fixing $c = c^{(v)}$ and $B = B^{(v)}$, refine the weights by
    $$w_{ij} = \frac{K_h(B^T f^{ij})}{\sum_{i = 1}^n K_h(B^T f^{ij})}$$
    and calculate the solutions of $(a_j, b_j), j = 1, \ldots, n$, to (\ref{4.2.1}):
    \begin{gather*}
    \begin{pmatrix}
    a_j^{(v)} \\
    b_j^{(v)} \\
    \end{pmatrix} = \left\{\sum_{i = 1}^n w_{ij}
    \begin{pmatrix}
    1 \\
    B^T f^{ij} \\
    \end{pmatrix}
    \begin{pmatrix}
    1 \\
    B^T f^{ij} \\
    \end{pmatrix}^T\right\}^{-1} \times \sum_{i = 1}^n w_{ij}
    \begin{pmatrix}
    1 \\
    B^T f^{ij} \\
    \end{pmatrix}y_i.
    \end{gather*} Set $v$ to be $v + 1$ and go to Step~1.
\item[Step~4.] Repeat Steps~1-3 until convergence. Our estimates, denoted by $\hat{a}, \hat{b}, \hat{f}$ and $\hat{B}$, are then based on the final values of $a^{(v)}, b^{(v)}, c^{(v)}$ and $B^{(v)}$.
\end{itemize}
In the same spirit as the MAVE procedure, we may iterate between Steps~2 and 3. For this reason, the above procedure is two-step iterative, and each cycle consists of a transformation step followed by a MAVE step. On the basis of our experience, however, iterations between Steps~2 and 3 cannot improve the result and are not necessary.

\begin{rema}
The computational intensive part is the transformation step in which the convergence rate of the Gauss-Jordan procedure is only linear. Nevertheless, the Gauss-Jordan procedure appears to be acceptably fast, and the convergence is usually obtained in 3-5 iterations. Since the run time of monotone-smoothing-transformed MAVE increases linearly with the number of predictors, its complexity is $O(p)$ plus the complexity of MAVE. Some simulation results regarding the computation time are given in the supplementary material.
\end{rema}

To determine the structural dimension $d$, we use the following criterion that was introduced for MAVE by Wang and Yin (2008):
$$\log\left(\frac{{\rm RSS}_k}{n}\right) + \frac{\log(n)}{nh^k} \times k,$$
where $k$ is the estimate of the dimension and
$${\rm RSS}_k = \sum_{j = 1}^n\sum_{i = 1}^n (y_i - \hat{a}_j - \hat{b}_j^T \hat{B}^T \hat{f}^{ij})^2 w_{ij}$$
is the residual sum of squares from the local linear smoothing. This criterion is similar in spirit to BIC.

\subsection{Examples}\label{sec43}

In this section we examine the finite-sample performance of monotone-smoothing-transformed MAVE. Our limited experience gained through simulation indicates that the method works quite well in terms of both subspace estimation and dimension determination. Four examples are considered. In each example, we generate 200 datasets with the sample size $n = 100$ and $n = 200$. The smoothing parameter is set to $\lambda = 0.001$. Let $\sigma_j$ denote the standard deviation of $f_j$.

\noindent {\sc Example 1.} Consider the following model
$$Y = \log\{(f_1 + f_2)^2 + 1\} \times (f_3 + f_4) + 0.5 \epsilon,$$ where $f = (f_1, \ldots, f_p) \sim N(0, \Sigma_f)$ with $(\Sigma_f)_{ij} = \rho^{|i - j|}$ for $1 \leq i, j \leq p = 6$, $\epsilon \sim N(0, 1)$, and $f$ and $\epsilon$ are independent. Two values of $\rho$ are explored, 0 and 0.5. In this example, $d_{Y|f(X)} = 2$ and $d_{Y|X} = 4$. $S_{Y|f(X)}$ is spanned by $\eta_1^f = (1, 1, 0, 0, 0, 0)^T$ and $\eta_2^f = (0, 0, 1, 1, 0, 0)^T$, while $S_{Y|X}$ is spanned by $e_1, e_2, e_3$ and $e_4$. To sample data from transformed Gaussian distributions, we first generate $f = f(X)$ from $N(0, \Sigma_f)$, then set $X_5 = f_5, X_6 = f_6$ and use probability integral transformation to generate $X_j$ for $j = 1, \ldots, 4$. Specifically, $X_1, X_2, X_3$ and $X_4$ have normal mixture distributions with respectively the skewed unimodal density \#2, the strongly skewed density \#3, the kurtotic unimodal density \#4 and the bimodal density \#6 used in Marron and Wand (1992)'s simulation study.

\noindent {\sc Example 2.} Let $f_1 = 2 \exp(X_1 / 3), f_2 = X_2, f_3 = X_3^3 / 3$ and $f_j = X_j$ for $j = 4, \ldots, p = 6$. The regression model has the form
$$Y =  f_1 + f_2^2 + f_3 + f_4 + f_5 + 0.5 \epsilon,$$
where $X \sim N(0, \Sigma_X)$ with $(\Sigma_X)_{ij} = \rho^{|i - j|}$ for $1 \leq i, j \leq 6$, $\epsilon \sim N(0, 1)$, and $X$ and $\epsilon$ are independent. Two values of $\rho$ are explored, 0 and 0.5. In this example, $d_{Y|f(X)} = 2$ and $d_{Y|X} = 4$. Further, $S_{Y|f(X)}$ is spanned by $\eta_1^f = (\sigma_1, 0, \sigma_3, \sigma_4, \sigma_5, 0)^T$ and $\eta_2^f = (0, 1, 0, 0, 0, 0)^T$, while $S_{Y|X}$ is spanned by $e_1, e_2, e_3$ and $\eta_4 = (0, 0, 0, 1, 1, 0)^T$.

\noindent {\sc Example 3.} We let $f_1 = 2 \exp(X_1 / 3), f_2 = X_2, f_3 = {\rm{sign}}(X_3) \times X_3^2 / 2$ and $f_j = X_j$ for $j = 4, \ldots, p = 6$. Consider the model
$$Y =  (f_1 + f_2) \times (f_3 + f_4 + f_5 + 1) + 0.5 \epsilon,$$
where $X \sim N(0, \Sigma_X)$ with $(\Sigma_X)_{ij} = \rho^{|i - j|}$ for $1 \leq i, j \leq 6$, $\epsilon \sim N(0, 1)$, and $X$ and $\epsilon$ are independent. Two values of $\rho$ are explored, 0 and 0.5. In this example, $d_{Y|f(X)} = 2$ and $d_{Y|X} = 4$. Further, $S_{Y|f(X)}$ is spanned by $\eta_1^f = (\sigma_1, \sigma_2, 0, 0, 0, 0)^T$ and $\eta_2^f = (0, 0, \sigma_3, \sigma_4, \sigma_5, 0)^T$, while $S_{Y|X}$ is spanned by $e_1, e_2, e_3$ and $\eta_4 = (0, 0, 0, 1, 1, 0)^T$.

\noindent {\sc Example 4.} Let $f_1 = X_1^3 / 3, f_2 = X_2, f_3 = 3\exp(2 X_3) / \{1 + \exp(2 X_3)\}$ and $f_j = X_j$ for $j = 4, \ldots, p = 6$. The regression model is of the form
$$Y = f_1 + (f_2 + f_3) \times (f_4 + f_5) + 0.3 \epsilon,$$
where $X \sim N(0, \Sigma_X)$ with $(\Sigma_X)_{ij} = \rho^{|i - j|}$ for $1 \leq i, j \leq 6$, $\epsilon \sim N(0, 1)$, and $X$ and $\epsilon$ are independent. Two values of $\rho$ are explored, 0 and 0.5. In this example, $d_{Y|f(X)} = 3$ and $d_{Y|X} = 4$. Further, $S_{Y|f(X)}$ is spanned by $\eta_1^f = (1, 0, 0, 0, 0, 0)^T, \eta_2^f = (0, \sigma_2, \sigma_3, 0, 0, 0)^T$ and $\eta_3^f = (0, 0, 0, \sigma_4, \sigma_5, 0)^T$, while $S_{Y|X}$ is spanned by $e_1, e_2, e_3$ and $\eta_4 = (0, 0, 0, 1, 1, 0)^T$.

For comparison, we also apply MAVE directly for the regression of $Y$ on $X$. As before, in the first part of the simulation study we assume that the structural dimension is known. The means and standard deviations of VCC and TCC, based on 200 repetitions, are summarized in Tables \ref{tab21} and \ref{tab22}. Since $d_{Y|X} = 4$, MAVE suffers from the curse of dimensionality. Within the proposed framework, however, the structural dimension is often greatly reduced; in this study $d_{Y|f(X)} = 2$ or $d_{Y|f(X)} = 3$. As we can see, the performance of monotone-smoothing-transformed MAVE, denoted by T-MAVE, is pretty well in all the examples considered here. Further, increasing the sample size generally improves the performance, and the results change little when we tune the correlation coefficient among the predictors.

Next, we study empirical aspects of the BIC-type criterion for determining the structural dimension. The empirical counts out of 200 repetitions are presented in Tables \ref{tab23} and \ref{tab24}. For MAVE, the BIC-type criterion tends to consistently underestimate the structural dimension in all four examples. However, the situation is different for T-MAVE. The numerical results, especially when the sample size is moderate ($n = 200$), indicate that for T-MAVE the BIC-type criterion should be consistent.

\begin{table}[htb!]\caption{The means and standard deviations (in parentheses) of the vector correlation coefficient (VCC) and the trace correlation coefficient (TCC), based on 200 repetitions, are reported for MAVE and transformed MAVE (T-MAVE) when $\rho = 0$} \label{tab21} \vspace{-0.3cm}
\centering
 {\small\scriptsize\hspace{12.5cm}
\renewcommand{\arraystretch}{1}\tabcolsep 0.15cm
\begin{tabular}{cccccccccc}
\hline
& & \multicolumn{2}{c}{Example 1} & \multicolumn{2}{c}{Example 2} & \multicolumn{2}{c}{Example 3} & \multicolumn{2}{c}{Example 4}\\
& & $n = 100$ & $n = 200$ & $n = 100$ & $n = 200$ & $n = 100$ & $n = 200$ & $n = 100$ & $n = 200$\\
MAVE  & VCC                            & 0.3781 & 0.4492 & 0.4716 & 0.5411 & 0.4942 & 0.6617 & 0.4710 & 0.6491  \\
      &                                &(0.2539)&(0.2713)&(0.2927)&(0.3045)&(0.2973)&(0.3027)&(0.3011)&(0.3229) \\
      & TCC                            & 0.8813 & 0.8967 & 0.9033 & 0.9176 & 0.9061 & 0.9375 & 0.9075 & 0.9372  \\
      &                                &(0.0411)&(0.0422)&(0.0445)&(0.0449)&(0.0465)&(0.0468)&(0.0427)&(0.0497) \\
T-MAVE& VCC                            & 0.9050 & 0.9858 & 0.9769 & 0.9866 & 0.9709 & 0.9789 & 0.9420 & 0.9935  \\
      &                                &(0.1709)&(0.0159)&(0.0149)&(0.0111)&(0.0148)&(0.0087)&(0.1310)&(0.0057) \\
      & TCC                            & 0.9578 & 0.9929 & 0.9884 & 0.9933 & 0.9854 & 0.9894 & 0.9833 & 0.9978  \\
      &                                &(0.0662)&(0.0079)&(0.0074)&(0.0055)&(0.0074)&(0.0043)&(0.0317)&(0.0018) \\
\hline
\end{tabular} }
\end{table}

\begin{table}[htb!]\caption{The means and standard deviations (in parentheses) of the vector correlation coefficient (VCC) and the trace correlation coefficient (TCC), based on 200 repetitions, are reported for MAVE and transformed MAVE (T-MAVE) when $\rho = 0.5$} \label{tab22} \vspace{-0.3cm}
\centering
 {\small\scriptsize\hspace{12.5cm}
\renewcommand{\arraystretch}{1}\tabcolsep 0.15cm
\begin{tabular}{cccccccccc}
\hline
& & \multicolumn{2}{c}{Example 1} & \multicolumn{2}{c}{Example 2} & \multicolumn{2}{c}{Example 3} & \multicolumn{2}{c}{Example 4}\\
& & $n = 100$ & $n = 200$ & $n = 100$ & $n = 200$ & $n = 100$ & $n = 200$ & $n = 100$ & $n = 200$\\
MAVE  & VCC                            & 0.4825 & 0.5721 & 0.4106 & 0.3737 & 0.4775 & 0.5790 & 0.5601 & 0.7615  \\
      &                                &(0.2638)&(0.2990)&(0.2652)&(0.2746)&(0.2688)&(0.3073)&(0.2903)&(0.2737) \\
      & TCC                            & 0.9002 & 0.9205 & 0.8922 & 0.8945 & 0.9007 & 0.9237 & 0.9184 & 0.9546  \\
      &                                &(0.0415)&(0.0459)&(0.0395)&(0.0351)&(0.0423)&(0.0458)&(0.0439)&(0.0452) \\
T-MAVE& VCC                            & 0.9088 & 0.9838 & 0.9685 & 0.9825 & 0.9630 & 0.9761 & 0.9131 & 0.9846  \\
      &                                &(0.1514)&(0.0133)&(0.0198)&(0.0157)&(0.0177)&(0.0096)&(0.1346)&(0.0199) \\
      & TCC                            & 0.9588 & 0.9919 & 0.9843 & 0.9912 & 0.9815 & 0.9880 & 0.9742 & 0.9949  \\
      &                                &(0.0560)&(0.0065)&(0.0098)&(0.0077)&(0.0088)&(0.0048)&(0.0310)&(0.0063) \\
\hline
\end{tabular} }
\end{table}

\begin{table}[htb!]\caption{The frequencies of estimated structural dimension out of 200 repetitions by the BIC-type criterion when $\rho = 0$. $d_{Y|f(X)} = 2$ in Examples 1, 2 and 3, and $d_{Y|f(X)} = 3$ in Example 4} \label{tab23} \vspace{-0.3cm}
\centering
 {\small\scriptsize\hspace{12.5cm}
\renewcommand{\arraystretch}{1}\tabcolsep 0.15cm
\begin{tabular}{cccccccccc}
\hline
& & \multicolumn{2}{c}{Example 1} & \multicolumn{2}{c}{Example 2} & \multicolumn{2}{c}{Example 3} & \multicolumn{2}{c}{Example 4}\\
& & $n = 100$ & $n = 200$ & $n = 100$ & $n = 200$ & $n = 100$ & $n = 200$ & $n = 100$ & $n = 200$\\
MAVE  & $\hat{d}_{Y|X} < 4      $      & 200    &200& 200    & 200    & 200    & 200    & 200    & 200     \\
      & $\hat{d}_{Y|X} = 4      $      & 0      &0  & 0      & 0      & 0      & 0      & 0      & 0       \\
      & $\hat{d}_{Y|X} > 4      $      & 0      &0  & 0      & 0      & 0      & 0      & 0      & 0       \\
T-MAVE& $\hat{d}_{Y|f(X)} < d_{Y|f(X)}$& 2      &2  & 0      & 0      & 0      & 0      & 89     & 1       \\
      & $\hat{d}_{Y|f(X)} = d_{Y|f(X)}$& 188    &194& 187    & 187    & 198    & 200    & 111    & 199     \\
      & $\hat{d}_{Y|f(X)} > d_{Y|f(X)}$& 10     &6  & 13     & 13     & 2      & 0      & 0      & 0       \\
\hline
\end{tabular} }
\end{table}

\begin{table}[htb!]\caption{The frequencies of estimated structural dimension out of 200 repetitions by the BIC-type criterion when $\rho = 0.5$. $d_{Y|f(X)} = 2$ in Examples 1, 2 and 3, and $d_{Y|f(X)} = 3$ in Example 4} \label{tab24} \vspace{-0.3cm}
\centering
 {\small\scriptsize\hspace{12.5cm}
\renewcommand{\arraystretch}{1}\tabcolsep 0.15cm
\begin{tabular}{cccccccccc}
\hline
& & \multicolumn{2}{c}{Example 1} & \multicolumn{2}{c}{Example 2} & \multicolumn{2}{c}{Example 3} & \multicolumn{2}{c}{Example 4}\\
& & $n = 100$ & $n = 200$ & $n = 100$ & $n = 200$ & $n = 100$ & $n = 200$ & $n = 100$ & $n = 200$\\
MAVE  & $\hat{d}_{Y|X} < 4      $      & 200    &200& 200    & 200    & 200    & 200    & 200    & 200     \\
      & $\hat{d}_{Y|X} = 4      $      & 0      &0  & 0      & 0      & 0      & 0      & 0      & 0       \\
      & $\hat{d}_{Y|X} > 4      $      & 0      &0  & 0      & 0      & 0      & 0      & 0      & 0       \\
T-MAVE& $\hat{d}_{Y|f(X)} < d_{Y|f(X)}$& 0      &0  & 0      & 0      & 0      & 0      & 135    & 36      \\
      & $\hat{d}_{Y|f(X)} = d_{Y|f(X)}$& 195    &198& 195    & 194    & 199    & 199    & 65     & 164     \\
      & $\hat{d}_{Y|f(X)} > d_{Y|f(X)}$& 5      &2  & 5      & 6      & 1      & 1      & 0      & 0       \\
\hline
\end{tabular} }
\end{table}

\section{Horse mussel data}

A sample of 82 horse mussels was collected in the Marlborough Sounds off the coast of New Zealand. The data were part of a larger ecological study of the mussels (Cook 1998; Cook and Weisberg 1999). The response variable is muscle mass $M$, the edible portion of the mussel, in grams. The four quantitative predictors are the height $H$, the length $L$, the width $W$ and the mass $S$ of the mussel's shell; $H$, $L$ and $W$ are in millimeters and $S$ is in grams. We are interested in studying the regression of $M$ on $(H, L, W, S)^T$.

The scatterplot matrix in the supplementary material shows that many of the predictor plots have approximately linear mean functions, but the mean functions for the plots including $S$ are clearly curved. Further, the (inverse) response plots for $H$, $L$ and $W$ show curved regression functions of roughly the same shape, while the (inverse) response plot for $S$ seems linear. Thus, the regression of $M$ on $(H, L, W, S)^T$ is evidently complicated with a greater than one-dimensional structure. We continue the analysis by replacing the predictors by their probability integral transformations and Yeo-Johnson transformations, respectively. As we can see from the scatterplot matrices in the supplementary material, the set of transformed predictors now satisfy the assumption of linearly related or normally distributed predictors, and the four (inverse) marginal response plots, which have the same shape, indicate a one-dimensional structure.

\begin{figure*}[!t]
\centering
\subfigure[] {\includegraphics[height=2in,width=2in,angle=0]{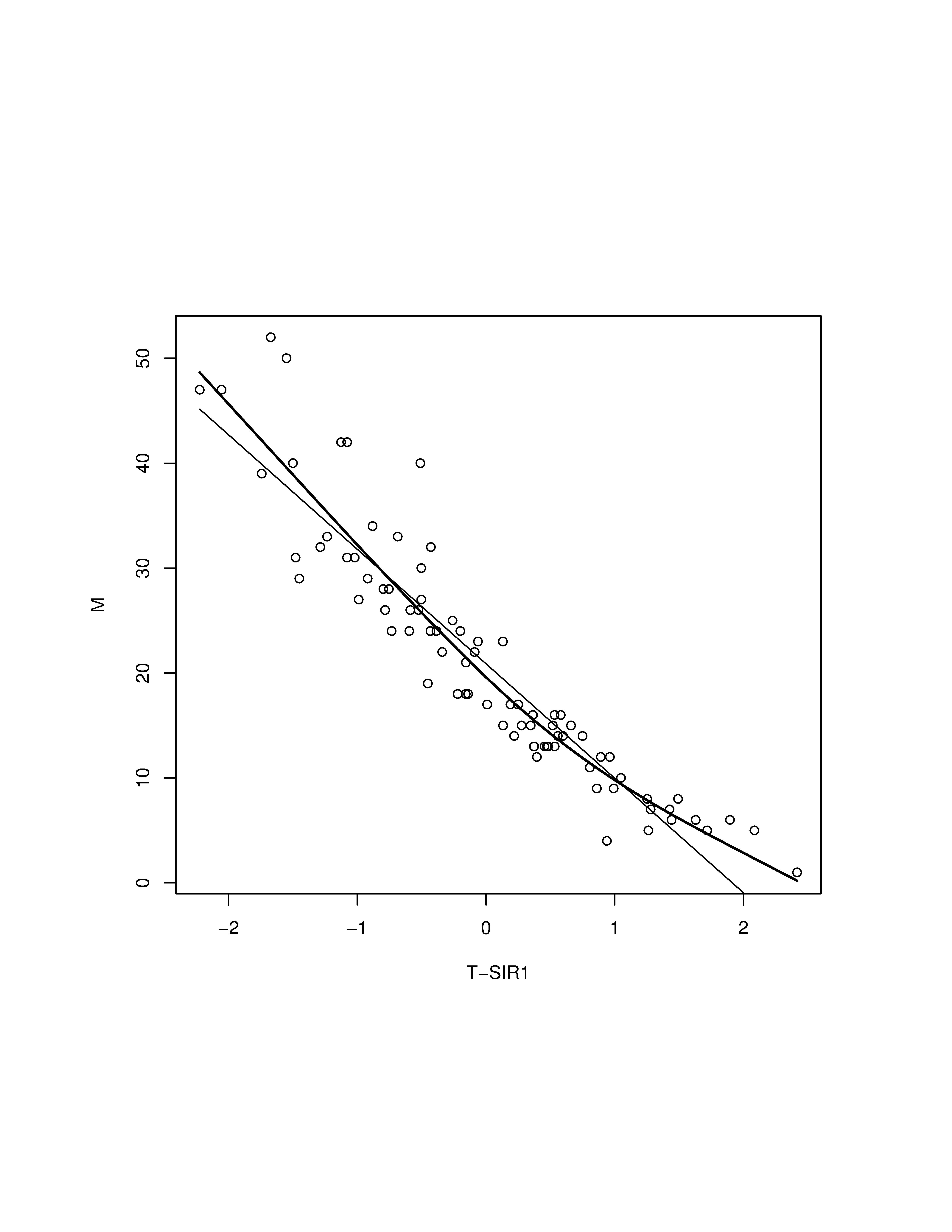}}
\subfigure[] {\includegraphics[height=2in,width=2in,angle=0]{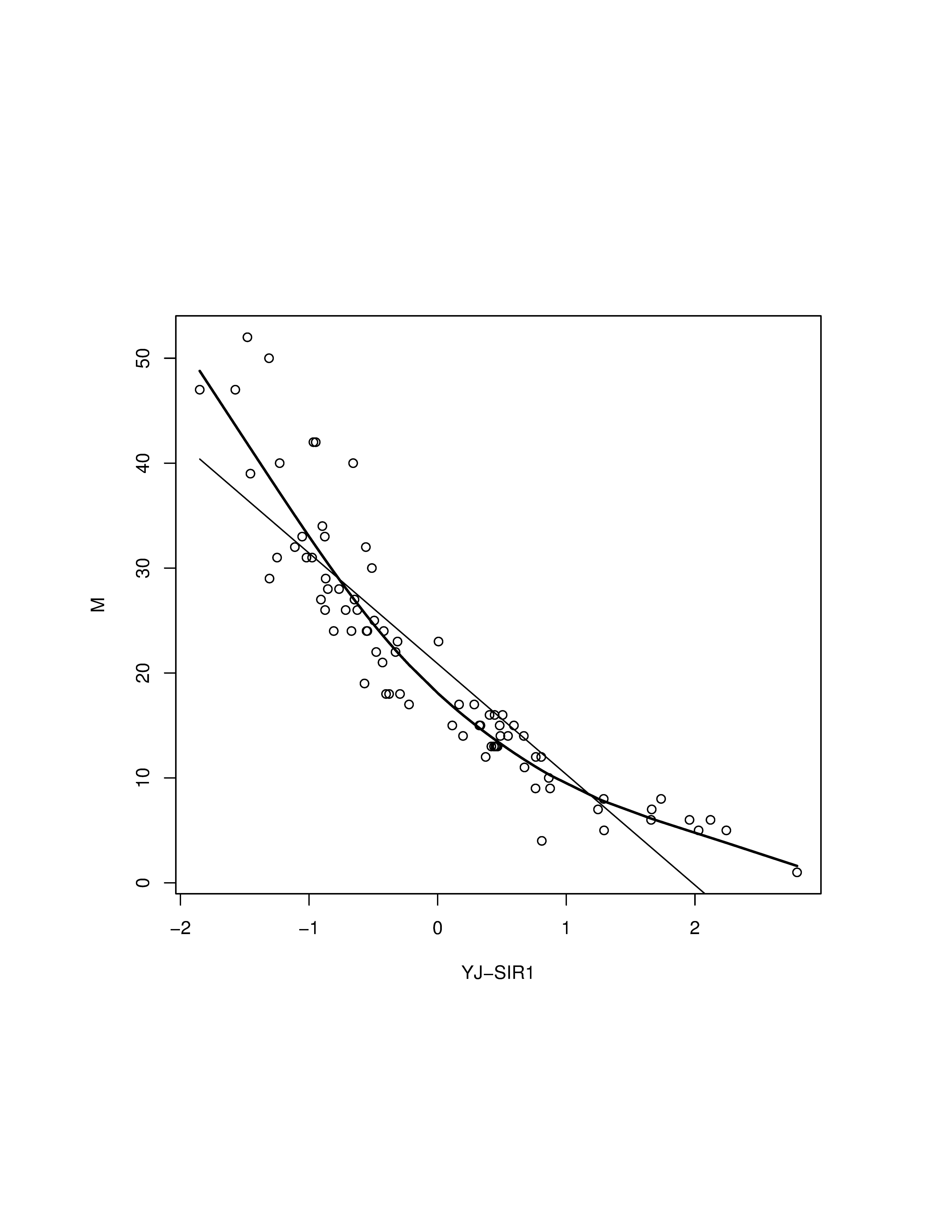}}
\captionsetup{width=1\textwidth}
\caption{Scatterplots of mussel mass $M$ versus the extracted predictor when five slices are used. The fitted curves are from linear regression (thin solid line) and smoothing spline (thick solid line). (a) transformed predictors using the probability integral transformation and (b) transformed predictors using the Yeo-Johnson transformation} \label{sirtsir}
\end{figure*}

\begin{figure*}[!t]
\centering
\subfigure {\includegraphics[height=4.5in,width=4.5in,angle=0]{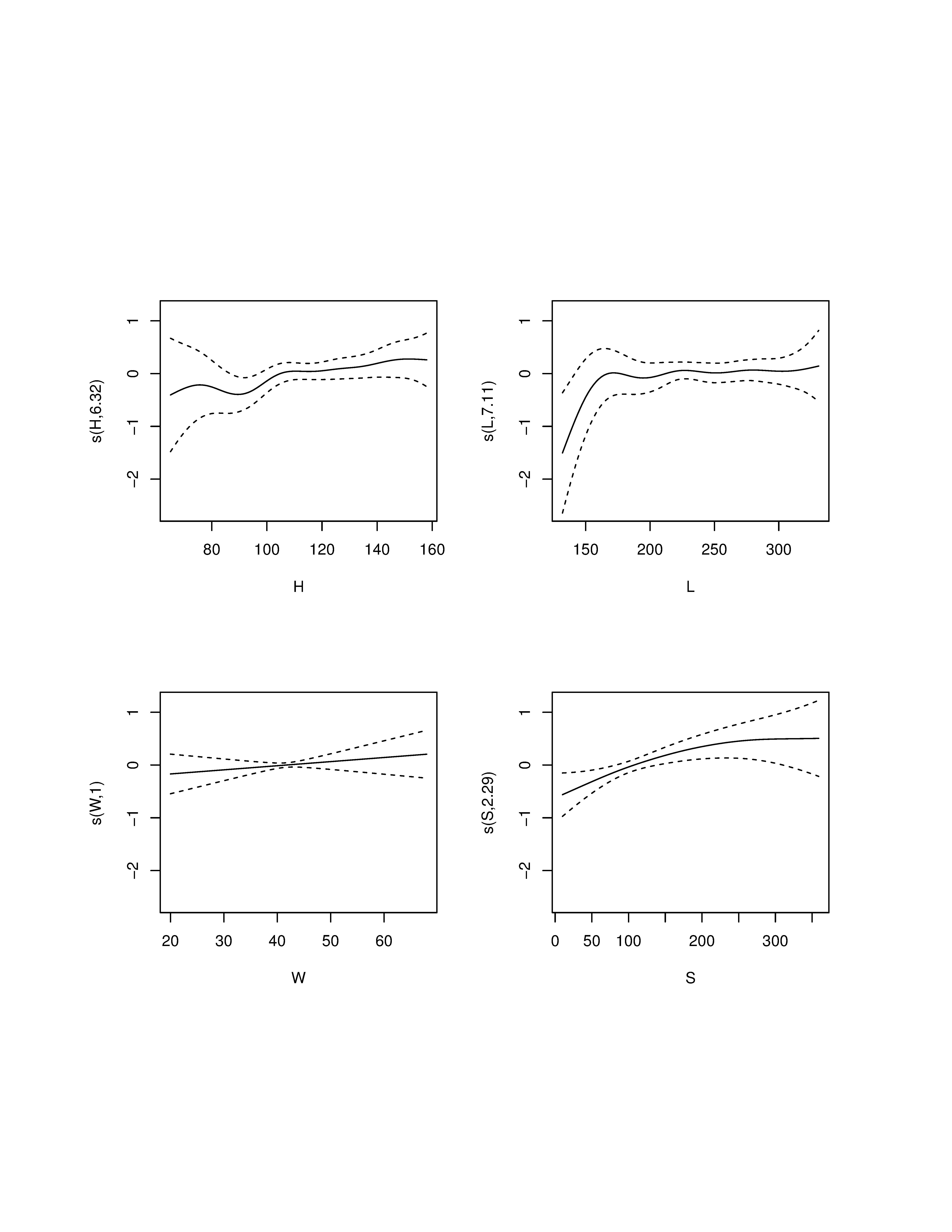}}
\captionsetup{width=1\textwidth}
\caption{Plots of the estimated terms of an additive model. The number in each $y$-axis caption is the effective degrees of freedom of the term being plotted. Solid curves are the function estimates and dashed curves delimit the 95\% confidence intervals for each function.} \label{gamterms}
\end{figure*}

Applying SIR, T-SIR and YJ-SIR, we find that both the sequential test and the BIC-type criterion are insensitive to the number of slices (five slices and ten slices) and give the same results: the estimated structural dimensions are 2, 1 and 1 respectively for SIR, T-SIR and YJ-SIR. Therefore, predictor transformations have the potential to reduce the structural dimension. Applying T-MAVE and the corresponding BIC-type criterion further confirms the one-dimensional structure. The scatterplots of mussel mass $M$ versus the extracted predictors, for T-SIR and YJ-SIR when five slices are used, are presented in Figure \ref{sirtsir}, and with the fitted parametric and nonparametric lines/curves superimposed. We see that there is evidence of an additive model after taking perhaps the log transformation of the response variable. To confirm this observation, we fit an additive model of $\log(M)$ on $(H, L, W, S)^T$. The $R$-squared and the adjusted $R$-squared are 93.8\% and 92.2\%, respectively. Figure \ref{gamterms} shows the estimated effects along with 95\% confidence intervals. We see that monotone predictor transformations are reasonable and, in particular, a monotone transformation of shell mass $S$ is desirable. 

\begin{thebibliography}{}

\item
Box, G. E. P. and Cox, D. R. (1964) An analysis of transformations (with discussion). {\it Journal of the Royal Statistical Society}, Ser. B, {\bf 26}, 211--252.

\item
Breiman, L. and Friedman, J. (1985) Estimating optimal transformations for multiple regression and correlation. {\it Journal of the American Statistical Association}, {\bf 80}, 580--598.


\item
Bura, E. and Cook, R. D. (2001) Extending sliced inverse regression: the weighted chi-square test. {\it Journal of the American
Statistical Association}, {\bf 96}, 996--1003.

\bibitem{Chen Li(1998)}
Chen, C. H. and Li, K. C. (1998) Can SIR be as popular as multiple linear regression? {\it Statistica Sinica}, {\bf 8}, 289--316.

\bibitem{Cook(1998)}
Cook, R. D. (1998) {\it Regression Graphics: Ideas for Studying
Regressions through Graphics}. {John Wiley \& Sons, New York}.

\item
Cook, R. D. (2007) Fisher lecture: Dimension reduction in regression. {\it Statistical Science}, {\bf 22}, 1--26.

\item
Cook, R. D. and Li, B. (2002) Dimension reduction for conditional mean in regression. \textit{The Annals of Statistics}, \textbf{30}, 455-474.


\item
Cook, R. D. and Ni, L. (2005) Sufficient dimension reduction via inverse regression: A minimum discrepancy approach. {\it Journal of the American Statistical Association}, {\bf 100}, 410--428.

\bibitem{Cook and Weisberg (1991)}
Cook, R. D. and  Weisberg, S. (1991) Comment on ``Sliced
inverse regression for dimension reduction". {\it Journal of the American Statistical Association}, {\bf 86}, 328--332.

\bibitem{Cook and Weisberg(1999)}
Cook, R. D. and  Weisberg, S. (1999) {\it Applied Regression
Including Computing and Graphics}. {John Wiley \& Sons, New York}.








\item
Li, B., Artemiou, A. and Li, L. (2011) Principle support vector machines for linear and nonlinear sufficient dimension reduction. {\it The Annals of Statistics}, {\bf 39}, 3182--3210.

\item
Li, B. and Dong, Y. (2009) Dimension reduction for nonelliptically
distributed predictors. {\it The Annals of Statistics}, {\bf 37}, 1272--1298.

\bibitem{Li and Wang (2007)}
Li, B. and Wang, S. (2007) On directional regression for dimension reduction. {\it Journal of the American Statistical Association}, {\bf 102}, 997--1008.

\item
Li, B., Wen, S. and Zhu, L. X. (2008) On a projective resampling method for dimension reduction
with multivariate responses. {\it Journal of the American Statistical Association}, {\bf 103}, 1177--1186.

\bibitem{Li (1991)}
Li, K. C. (1991) Sliced inverse regression for dimension reduction. {\it Journal of the American Statistical Association}, {\bf 86}, 316--327.

\bibitem{Li (1992)}
Li, K. C. (1992) On principal Hessian directions for data visualization and dimension reduction: Another application of Stein's lemma. {\it Journal of the American Statistical Association}, {\bf 87}, 1025--1039.


%

\item
Ma, Y. and Zhu, L. P. (2012) A semiparametric approach to dimension reduction. {\it Journal of the American Statistical Association}, {\bf 107}, 168--179.

\item
Marron, J. S. and Wand, M. P. (1992) Exact mean integrated squared error. {\it The Annals of Statistics}, {\bf 20}, 712--736.


\item
Ramsay, J. O. (1988) Monotone regression splines in action (with discussion). {\it Statistical Science}, {\bf 3}, 425--461.

\item
Ramsay, J. O. (1998) Estimating smooth monotone functions. {\it Journal of the Royal Statistical Society}, Ser. B, {\bf 60}, 365--375.

\item
Ramsay, J. O. and Silverman, B. W. (2005) {\it Functional Data Analysis}. Springer, New York.



\item
Wu, H. M. (2008) Kernel sliced inverse regression with applications to classification. {\it Journal of Computational and Graphical Statistics}, {\bf 17}, 590--610.

\item
Wu, Q., Liang, F. and Mukherjee, S. (2007) Regularized sliced inverse regression for kernel
models. Technical Report 07-25, ISDS, Duke University.

\item
Xia, Y., Tong, H., Li, W. K. and Zhu, L. X. (2002) An adaptive estimation of dimension reduction space. {\it Journal of the Royal Statistical Society}, Ser. B, {\bf 64}, 363--410.

\item
Yeo, I. and Johnson, R. A. (2000) A new family of power transformations to improve normality or symmetry. {\it Biometrika}, {\bf 87}, 954--959.

\item
Yin, X. and Li, B. (2011) Sufficient dimension reduction based on an ensemble of minimum average variance estimators. \textit{The Annals of Statistics}, \textbf{39}, 3392-3416.

\item
Yin, X., Li, B. and Cook, R. D. (2008) Successive direction extraction for estimating the central subspace in a multiple-index regression. {\it Journal of Multivariate Analysis}, {\bf 99}, 1733--1757.

\bibitem{Zhu et al. (2009)}
Zhu, L. P., Wang, T., Zhu, L. X. and Ferr\'{e}, L. (2010) Sufficient dimension reduction through discretization-expectation estimation. {\it Biometrika}, {\bf 97}, 295--304.

\item
Zhu, L. X., Miao, B. and Peng, H. (2006) On sliced inverse regression with high-dimensional covariates. {\it Journal of the American Statistical Association}, {\bf 101}, 630--643.


\end{thebibliography}
\end{document}